\documentclass[aps,prl,12pt,superscriptaddress,longbibliography,nourl,nodoi,noeprint,onecolumn]{revtex4-2}
\usepackage{amsmath,amssymb,physics,bm}
\usepackage{xcolor,soul,setspace}
\usepackage{graphicx,microtype,siunitx} 
\usepackage[colorlinks=true,linkcolor=blue,citecolor=blue,urlcolor=blue]{hyperref} 
\usepackage[all]{hypcap} 

\newcommand{\kB}[0]{k_{\rm B}}
\newcommand{\rvec}[0]{{\bm r}} 
\newcommand{\nvec}[0]{{\bm n}}
\newcommand{\ua}[0]{v_{\rm a}} 
\newcommand{\Dr}[0]{D_\mathrm{r}}
\newcommand{\Dc}[0]{D_{\rm c}} 
\newcommand{\muc}[0]{\mu_{\rm c}}
\newcommand{\RH}[0]{R_{\rm h}}

\newcommand{\Deff}[0]{\tilde{D}} 
\newcommand{\mueff}[0]{\tilde{\mu}}
\newcommand{\xic}[0]{\xi_{\rm c}}
\newcommand{\xir}[0]{\xi_{\rm r}}
\newcommand{\ee}[0]{\mathrm{e}} 

\begin{document} 
\title[]{Nanoswimmers in a ratchet potential: Effects of a transverse rocking force}

\author{Mykola Tasinkevych}
\email[]{mykola.tasinkevych@ntu.ac.uk}
\affiliation{Centro de F\'{i}sica Te\'{o}rica e Computacional, Faculdade de Ci\^{e}ncias, Universidade de Lisboa, 1749-016 Lisboa, Portugal} 
\affiliation{Departamento de F\'{\i}sica, Faculdade de Ci\^{e}ncias, Universidade de Lisboa, 1749-016 Lisboa, Portugal}
\affiliation{SOFT Group, School of Science and Technology, Nottingham Trent University, Clifton Lane, Nottingham NG11~8NS, United Kingdom}
\affiliation{International Institute for Sustainability with Knotted Chiral Meta Matter, Hiroshima University, Higashihiroshima 739-8511, Japan}

\author{Artem Ryabov}
\email[]{rjabov.a@gmail.com}
\affiliation{Charles University, Faculty of Mathematics and Physics, Department of Macromolecular Physics, V Hole\v{s}ovi\v{c}k\'ach 2, CZ-18000 Praha 8, Czech Republic}

\date{October 29, 2023}

\setstretch{1.5}
\begin{abstract}
We study the dynamics of a chemical nanoswimmer in a ratchet potential, which is periodically rocked in the transverse direction. As a result of the mechanochemical coupling, the self-propulsion velocity becomes force-dependent and particle trajectories are rectified in the direction of the ratchet modulation. The magnitude and direction of the nanoswimmer mean velocity depend upon both the rocking amplitude and the frequency. Remarkably, for frequencies larger than the inverse correlation time of the rotational diffusion, the velocity exhibits oscillatory behaviour as a function of the amplitude and the frequency with multiple reversals of the sign. These findings suggest that mechanochemical coupling can be utilized for controlling the motion of chemically active particles at the nanoscale. 
\end{abstract}
\maketitle  
\section{Introduction}

Microscopic reversibility (MR)~\cite{Tolman:PNAS1925, Onsager:1931a, Astumian2012} is a fundamental concept in stochastic Markovian models of molecular motors operating at the nanoscale far from thermodynamic equilibrium~\cite{Astumian2019, Astumian/etal/2020, Feng/etal/2021}. The MR is a manifestation of the time-reversal symmetry of underlying microscopic equations of motion, which on the mesoscopic scale translates to stringent conditions on transition rates or/and on the form of possible Langevin equations. For Markov jump processes and diffusions when the motor is exchanging energy with weakly coupled equilibrium baths, MR leads to the local detailed balance condition~\cite{maes2021}.

An important consequence of the MR in the context of molecular machines is a phenomenon of mechanochemical coupling \cite{Zhang/Hess:2021, Amano/etal:2022} when the reaction and transition rates can be controlled by coupling motors to both the reservoir of chemical free energy and the reservoir of mechanical work~\cite{Astumian/Bier:1996, Carter2005, Sakamoto2008, Lipowsky/Liepelt:2008}. The mechanochemical coupling underlies the ability of molecular motors to exert forces and torques \cite{Toyabe/etal:2011, Mukherjee/etal:2015, Trivedi/etal:2020, Feng/etal:2021, Borsley/etal:2021}, and is also utilised in different single-molecule force spectroscopies~\cite{Bustamante/etal:2003, Bustamante/etal:2021, Volpe/etal:2022}, and for single-molecule sensing \cite{Hu/etal:2022}. 

Physical effects stemming from the mechanochemical coupling in a minimal thermodynamically consistent stochastic model of a chemically active nanoswimmer were recently analyzed theoretically in Refs.~\cite{Ryabov/Tasinkevych:SoftMatt2022, Ryabov/Tasinkevych:JCP2022, ryabov/MR:2022}. In particular, due to the mechanochemical coupling, the self-propulsion speed of the nanoswimmer becomes dependent on external forces. For a space-independent force, this results in an enhancement of the long-time diffusion constant~\cite{Ryabov/Tasinkevych:SoftMatt2022, Ryabov/Tasinkevych:JCP2022} over the one observed for a force-free motion. 

Moreover, in Ref.~\cite{ryabov/MR:2022}, this force-dependence of the active speed has been employed to design a ratchet effect capable of rectifying trajectories of the nanoswimmer diffusing in an asymmetric sawtooth-like potential. A distinctive feature of this ratchet effect is the rectification of the nanoparticle motion induced by a constant force applied in a direction orthogonal to that of the asymmetric potential modulation. While the model of~\cite{ryabov/MR:2022} was simple enough to allow us to understand the physical origins of this rectification, it suffers from a drawback, which can become an obstacle in practical applications: For a constant transverse force, there exists a mean net drift of nanoswimmers in the force direction. In an experiment, this would lead to the removal of nanoparticles from the observation area of a microscope or to the accumulation of nanoparticles at one of the system's boundaries. 

In the present article, we address this issue by considering a time-periodic transverse force with an average value equal to zero. Remarkably, such a generalisation of the original model of Ref.~\cite{ryabov/MR:2022}, gives rise to several unexpected effects. The most exciting of them is the oscillatory behaviour and multiple reversals of the nanoswimmer mean velocity induced by changing the amplitude of the transverse force, provided the driving frequency is above a certain threshold. These current reversals arise due to the mechanochemical coupling and will be absent in the active Brownian particle model with a constant active speed. Also, the current direction sensitively depends upon the reaction rates, which provides an opportunity for the development of efficient approaches for sorting and guiding chemical nanoswimmers.

\section{Active Brownian particle model with mechanochemical coupling}

We consider a nanoswimmer whose active (self-propelled) motion through the ambient fluid environment is powered by catalytic chemical reactions. In addition, the nanoswimmer undergoes rotational and translational diffusion and it is being acted upon by an external time-dependent force ${\bm F}(\rvec,t)$. 

The dynamics of the centre of mass position $\rvec(t) = (x(t),y(t))$ of such a nanoparticle can be described by the Langevin equation \cite{Pietzonka/Seifert:2018, Speck:2018, Speck:2019, Ryabov/Tasinkevych:SoftMatt2022}
\begin{equation} 
\label{eq:Langevin-general}
\frac{\dd \rvec }{\dd  t} = \ua (\rvec,t) \nvec(t) + \mu {\bm F}(\rvec,t) + \sqrt{2 D }\, {\bm \xi}(t).
\end{equation} 
Here, $[\mu {\bm F}(\rvec,t) + \sqrt{2 D }\, {\bm \xi}(t)]$
is the velocity of the overdamped (passive) Brownian motion with diffusion coefficient $D$ related to mobility $\mu$ and temperature $T$ of the ambient fluid via the fluctuation-dissipation theorem $D=\mu \kB T$, where $\kB$ is the Boltzmann constant; 
${\bm \xi}(t) = (\xi_x(t), \xi_y(t))$ denotes a zero-mean Gaussian white noise process. 

The term $\ua (\rvec,t) \nvec(t)$ on the right-hand side of Eq.~\eqref{eq:Langevin-general} represents the active velocity with magnitude $\ua (\rvec,t)$ and the direction $\nvec(t)$. In two dimensions, the rotational diffusion of $\nvec(t)=( \cos\phi(t) , \sin\phi(t) )$ is conveniently described by considering the Brownian motion of angle $\phi(t)$, i.e.,  
$\dot \phi(t) = \sqrt{2 \Dr }\, \xir(t)$, 
with $\Dr$ being the rotational diffusion coefficient, $\xir(t)$ a zero-mean Gaussian white noise. For a wide class of chemical reactions (outlined below), the active speed $\ua (\rvec,t)$ can be approximated by the expression \cite{Ryabov/Tasinkevych:SoftMatt2022}  
\begin{equation} 
\label{eq:ua}
\ua(\rvec,t) = u + \muc\, \nvec (t) \cdot {\bm F}(\rvec,t) + \sqrt{2 \Dc}\, \xic(t), 
\end{equation} 
where $u$ is the constant mean active speed in ${\bm F}=0$ case, and $\xic(t)$ is a Gaussian white noise. Positive constants $\Dc$ and $\muc$ are related by $\Dc = \muc \kB T $, and the dot $\cdot$ denotes the scalar product. Thus, in total, Langevin equation~\eqref{eq:Langevin-general} contains four white noise processes $\xi_\alpha(t)$, 
$\alpha, \beta \in \{x,y,\textrm{r},\textrm{c}\}$. These processes are statistically independent, they satisfy 
$\langle \xi_\alpha(t) \rangle =0$, and 
$\langle \xi_\alpha(t) \xi_\beta(t') \rangle = \delta_{\alpha\beta} \delta(t-t')$. 

To gain an insight into the origins of individual terms in Eq.~\eqref{eq:ua}, a more detailed (microscopic) level of description of the active chemically powered dynamics of the nanoswimmer is needed. In fact, Eq.~\eqref{eq:ua}
has been derived~\cite{Ryabov/Tasinkevych:SoftMatt2022} as a diffusion approximation to a microscopic jump-diffusion process, where the active speed $\ua (\rvec,t)$ is a random shot-noise-like process. Namely, at this microscopic level of description, chemically-powered displacements of the nanoswimmer are modelled by a Markov jump process with the forward ($k_+$, corresponding to jump $\rvec\to\rvec+\nvec \delta r$) and backward ($k_-$, yielding $\rvec\to\rvec-\nvec \delta r$) jump rates satisfying the local detailed balance condition~\cite{Maes:SciPost2021}
\begin{equation}
\frac{k_+}{k_-} = 
\ee^{( \Delta G_{\rm r} - \delta W) / \kB T }. 
\label{eq:detailed_balance}
\end{equation} 
These jumps of the nanoswimmer are caused by a reversible chemical reaction that dissipates free energy $\Delta G_{\rm r}$. Additionally, in each jump the work $\delta W \approx -\nvec \cdot {\bm F} \delta r$ is done against the external force ${\bm F}$. Eq.~\eqref{eq:ua} is then obtained as the diffusion approximation to such a microscopically reversible Markovian jump-diffusion process. 

In Eq.~\eqref{eq:ua}, the noise $\xic(t)$ causes fluctuations in the active speed and arises due to fluctuations in the number of forward and backward chemical reactions. 

A remarkable implication of the local detailed balance condition~\eqref{eq:detailed_balance} is that magnitudes of the reaction rates $k_\pm$ can be adjusted by external forces acting upon the nanoswimmer. In general, such a phenomenon is known as the mechanochemical (or chemomechanical) coupling~\cite{Beyer/Clausen-Schaumann:2005, Zhang/Hess:2021}. In the present model, the force-induced changes in the reaction rates give rise to the force-induced variation of the nanoswimmer mean speed.  Within the diffusion approximation, the effect of this mechanochemical coupling is captured in the expression~\eqref{eq:ua} for the active speed $\ua(\rvec,t)$, which depends on the scalar product $\nvec \cdot {\bm F}$, i.e., on the component of the total external force ${\bm F}$ in the direction $\nvec$ of the active motion.

\section{Transverse ratchet effect based on the mechanochemical coupling}

In our previous works~\cite{Ryabov/Tasinkevych:SoftMatt2022, Ryabov/Tasinkevych:JCP2022, ryabov/MR:2022} we demonstrated that the term  $\nvec \cdot {\bm F}$  in Eq.~\eqref{eq:ua} can have significant impacts on the particle dynamics. In the simplest case of a constant force of magnitude $F$ acting upon the nanoswimmer, its long-time effective diffusion constant becomes enhanced by $(\muc F)^2/32\Dr$ \cite{Ryabov/Tasinkevych:SoftMatt2022, Ryabov/Tasinkevych:JCP2022}. As a result, the nanoparticle can be expected to diffuse faster in those spatial regions where the force is stronger. 

We explored this force-induced control of diffusion further in Ref.~\cite{ryabov/MR:2022}, where we reported a rectification of the particle motion in a one-dimensional asymmetric $\lambda$-periodic potential 
\begin{equation} 
\label{eq:V}
V(x) = V_0 \left[ \sin\left(\frac{2\pi x}{\lambda} \right) + \frac{1}{4} \sin\left(\frac{4\pi x}{\lambda}\right) \right], 
\end{equation}  
with the barrier height $V_0$. Similar periodic asymmetric saw-tooth-like potentials have been used to rectify the particle motion in various theoretical and experimental studies~\cite{Reimann:2002, Angelani/etal:2011, Ai/Wu:2014, McDermott/etal:2016, Ryabov/etal:2016, Arzola/etal:2011, Lozano/etal:2016, Arzola/etal:2017, Skaug/etal:2018, Schwemmer/eta:2018, Stoop/etal:2019, Paneru/etal:2021, Leyva/etal:2022}. Typically in such studies, the rectification is achieved by a time-dependent force acting along the direction of the asymmetric potential modulation, i.e., in the $\hat{x}$-direction. Contrary, in Ref.~\cite{ryabov/MR:2022}, the net nanoparticle motion in the $-\hat{x}$-direction was triggered by applying a constant force $f_y$ in the perpendicular $\hat{y}$-direction. It was found that the nanoparticle's average speed has a global minimum, depending on $V_0$,  as a function of $f_y$. 

Here we discuss such a transverse ratchet effect for the case when the nanoswimmer is being acted upon by a time-periodic transverse force $f_y \cos(\Omega t)$. The total external force then reads  
\begin{equation}
\label{eq:F}
\bm F(\rvec , t) = \left( -\frac{\dd V}{\dd x}, f_y \cos(\Omega t)\right), 
\end{equation} 
and the Langevin equations~\eqref{eq:Langevin-general} become 
\begin{align}
\label{eq:Langevin-x}
& \frac{\dd x}{\dd  t} = \ua (\rvec,t) \cos\phi(t) - \mu \frac{\dd V}{\dd x} + \sqrt{2 D }\,\xi_x(t), \\ 
& \frac{\dd y}{\dd  t} = \ua (\rvec,t)\sin\phi(t) + \mu f_y \cos(\Omega t) + \sqrt{2 D }\,\xi_y(t),
\end{align} 
where 
\begin{equation}
\label{eq:ua-explicit}
\ua(\rvec,t) = u + \muc\! \left[  -\frac{\dd V}{\dd x} \cos\phi(t) + f_y \cos(\Omega t) \sin\phi(t) \right] + \sqrt{2 \Dc}\, \xic(t). 
\end{equation}
From these equations, we can see that the mechanochemical coupling term in $\ua(\rvec,t)$, i.e., the square bracket multiplied by $\muc$ in~\eqref{eq:ua-explicit}, induces time-dependent changes of the active speed. This translates into the time-dependent driving for the $x$-coordinate in~\eqref{eq:Langevin-x}. The magnitude of such driving is further modulated by a term dependent on the orientation of the particle, which changes by the rotational diffusion, i.e., by the Brownian motion of the angle $\phi(t)$. 

\begin{figure}[b!]
\centering
\includegraphics[width=0.5\columnwidth]{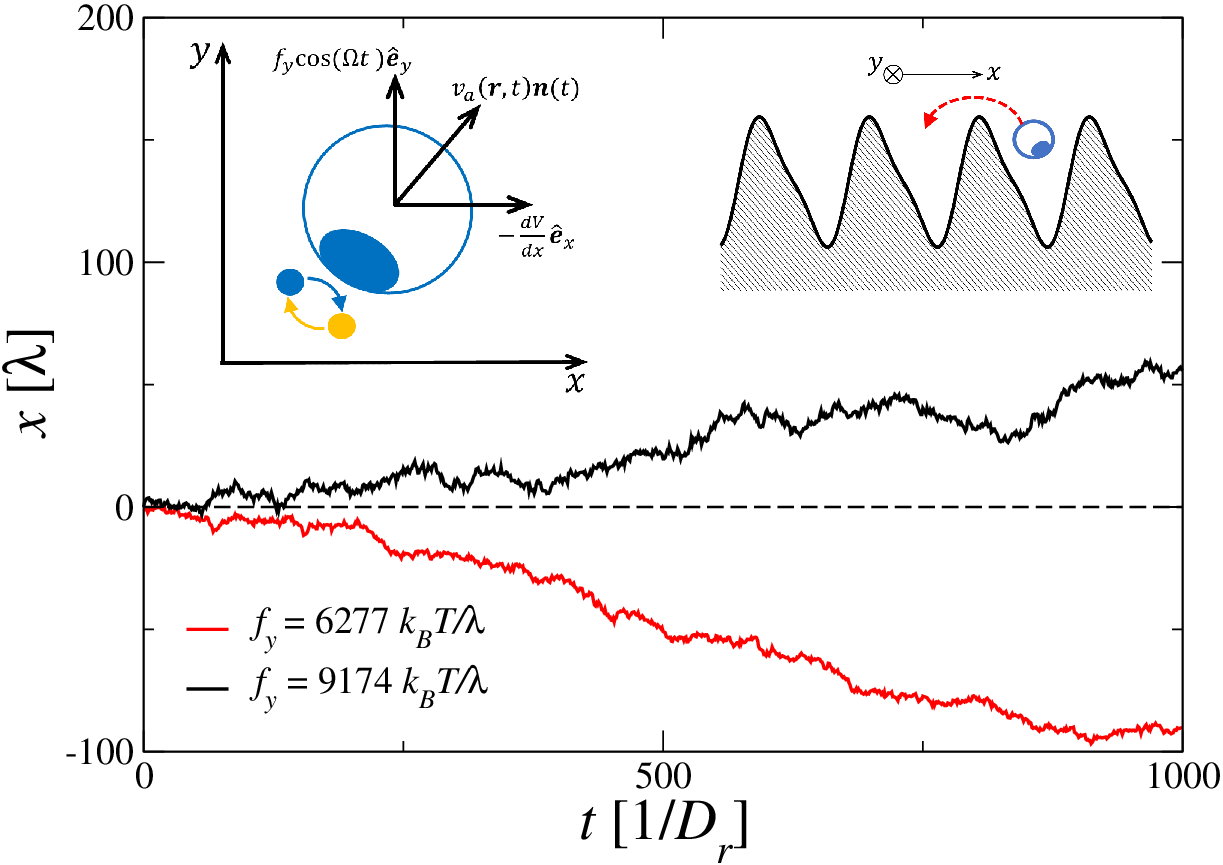} 
\caption{Typical trajectories $x(t)$ of the active nanoparticle self-propelling with the speed $\ua(\rvec ,t)$, Eq.~\eqref{eq:ua}, in the spatio-temporal force field given by Eq.~\eqref{eq:F} with $\Omega = 14\Dr$. The left inset illustrates schematically a chemically active nanoparticle, shown by a large open circle, moving with the speed $\ua(\rvec ,t)$ in the direction of $\nvec(t)$. Self-propulsion is powered by chemical reactions (blue and orange arrows), the small solid circles represent the reactant and the product particles, and the blue ellipse on the nanoparticle stands for a catalytic region. Assuming the microscopic reversibility of the self-propulsion mechanism gives rise to the mechanochemical coupling term $\muc \nvec \cdot \bm F$ in Eq.~\eqref{eq:ua} for $\ua(\rvec ,t)$. The $x$-component of the external force $-\dd V/\dd x$ is represented by the horizontal arrow, where ratchet potential $V(x)$ is given in Eq.~\eqref{eq:V}. The $y$-component of the external force $f_y\cos(\Omega t)$ is shown by the vertical arrow. For certain values of $\Omega$, the direction of the nanoparticle motion can be reversed by increasing the driving amplitude $f_y$. The right inset represents schematically the nanoswimmer moving in ratchet potential $V(x)$. The nanoswimmer drifts along the $x-$axis under the action of transverse force $f_y\cos(\Omega t)$ and the direction of motion, towards the left or right, depends on $f_y$, provided $\Omega$ is large enough. The physical mechanism of this ratchet effect is based on mechanochemical coupling.}
\label{fig:model}
\end{figure} 

While the time-periodic transverse force $f_y \cos(\Omega t)$ induces no net particle motion in the $\hat{y}$-direction, a non-zero mean particle velocity along the $x$-axis can arise due to the time-variation of $\ua(\rvec,t)$. As compared to the previously studied case with a constant transverse forcing~\cite{ryabov/MR:2022}, this velocity exhibits a rather complex behaviour. Below we report that for the frequency $\Omega$ above a certain threshold, the average nanoparticle velocity can be reversed by changing the amplitude $f_y$. On the level of individual trajectories, this fact is illustrated in Fig.~\ref{fig:model}, where the nanoparticle is subjected to a relatively fast periodic transverse driving. It then moves on average in the $-\hat{x}$-direction for small driving amplitudes $f_y$ (the red curve in Fig.~\ref{fig:model}) and in the $\hat{x}$-direction for larger $f_y$ (the black curve in Fig.~\ref{fig:model}).

Before embarking on the discussion of the results, let us note, that the reported transverse ratchet effect is not expected to occur in models of active particles that do not account for the mechanochemical coupling term in~\eqref{eq:ua-explicit}, e.g., in the paradigmatic active Brownian particle model~\cite{Howse/etal:2007, tenHagen/etal:2011, Zottl/Stark:JPCM2016} with constant active speed $\ua = u$. An exception can be a model of an asymmetric active Brownian particle having ellipsoidal shape~\cite{Kurzthaler/etal:2016}, see also equations of motions in Ref.~\cite{Grima/Yaliraki:2007} for the passive variant of the model. If we interpret $\bm F(\rvec, t)$ as the total force acting on a centre of mass of the ellipsoidal particle, equations of motion for the particle centre of mass in the current model become remarkably similar to those for the ellipsoidal particle. Hence at first glance, one can expect the transverse force to induce a rectification of active Brownian ellipsoids too. However, the dynamics of the ellipsoidal particle can be different than that in the current model due to the torque term present in the diffusion equation for the particle orientation. This torque term can cause significant dissimilarities in the results obtained for the two models.

In the following, all reported numerical results will be presented in the natural units: we choose $1/\Dr$ as the unit of time, $\lambda$ as the unit of length, and $\kB T$ of energy.

\section{Results and discussion}

We have integrated Langevin equations~\eqref{eq:Langevin-general}, with the self-propulsion speed \eqref{eq:ua} numerically employing the Euler–Maruyama method \cite{Kloeden/Platen:1992}. The values of the diffusion constants $D$ and $\Dr$ are obtained by using the Stokes-Einstein equation $D = \kB T/6\pi\eta \RH$, and $\Dr = \kB T/8\pi\eta \RH^3$ valid for a sphere of radius $\RH$, where $\eta$ is the fluid dynamic viscosity. With $\RH=15\;\si{nm}$ which is typical for size of enzymes \cite{Jee/etal:PNAS2018, Ah-Young/etal:2018, Jee:2020}, and setting $T = 300\;\si{K}$, $\eta = 8.53\times  10^{-4}\;\si{Ns/m^2}$, which is the dynamic viscosity of water at this temperature, we obtain $D \approx 1.7\times 10^{-11}\;\si{m^2/s}$ and $\Dr \approx 5.7\times 10^{4}\;\si{s^{-1}}$. The phenomenological parameter $\Dc$ is approximated as $\Dc \approx (\delta r)^2 (k_++k_-)/2$ \cite{Ryabov/Tasinkevych:SoftMatt2022}. We neglect the reverse reaction setting $k_-=0$, and by using $k_+ = 10^5\;\si{s^{-1}}$, $\delta r = 5\;\si{nm}$ we find $\Dc\approx 1.3\times 10^{-12}\;\si{m^2/s}$. We emphasize, that the reported ratchet effect relies only on the condition $\Dc >0$ and does not depend on the actual value of $\Dc$. Finally, we set the constant part of the self-propulsion speed~\eqref{eq:ua} $u=10\;\si{nm/s}$, but the reported results do not depend on the actual value of $u$ provided $u \lesssim 10^3\;\si{nm/s}$ \cite{ryabov/MR:2022}. The speed of $10^3\;\si{nm/s}$ corresponds to $\approx 33$ body lengths per second, which is a moderate value compared to some bacteria which can reach speeds of over 200 body lengths per second \cite{bente:2020}. 

We shall discuss the mean nanoparticle velocity $\langle v_x \rangle$ in the $\hat{x}$-direction calculated as 
\begin{equation}
\langle v_x \rangle = \frac{1}{N}\sum_{i=1}^{N} \frac{x_i(t_{\rm max})}{t_{\rm max}},
\end{equation} 
where $N$ is the number of realisations, $x_i(t_{\rm max})$ is the $x$-coordinate of the nanoparticle in realisation $i$ at final time $t_{\rm max}$. In simulations, we have used $\Dr t_{\rm max}=10^3$ and $N=10^4$.

\begin{figure}[t]
\centering
\includegraphics[width=0.5\columnwidth]{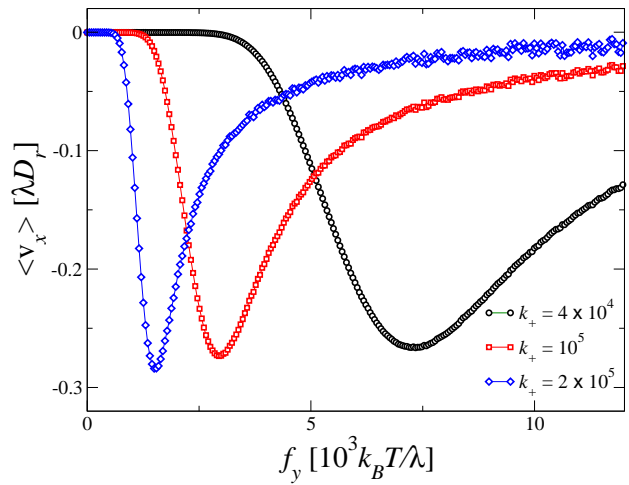}
\caption{The $x-$component of the particle mean velocity $\langle v_x \rangle$, as a function of the magnitude $f_y$ of a constant force, $\Omega=0$, applied perpendicular to the ratchet direction. The depth of the ratchet potential $V_0 = 10\; \kB T$. Different curves correspond to different values of the forward reaction rate constant $k_+$.}
\label{fig:vx_vs_fy_W0}
\end{figure}

Figure~\ref{fig:vx_vs_fy_W0} demonstrates the effect of $\Dc$  (through $k_+$) on $\langle v_x \rangle$, assuming initially constant transverse driving, i.e. $\Omega=0$. $\langle v_x \rangle$ as a function of $f_y$ reveals minimum, whose depth and location increases with increasing $k_+$. A qualitative understanding of the underlying mechanism for the rectification of the nanoparticle trajectory can be gained by approximating the Langevin equation \eqref{eq:Langevin-general} for the $x$-coordinate as follows 
\begin{equation} 
\label{eq:Langevin-x-approx}
\frac{\dd x}{\dd t} \approx  
\frac{\muc}{2}\sin\left(2\phi(t)\right)f_y\cos(\Omega t) 
- \mueff \frac{\dd V}{\dd x} 
+ \sqrt{2\Deff}\, \xi_x(t),  
\end{equation} 
where $\mueff = \mu +\muc/2$, and the enhanced diffusivity $\Deff = D+\Dc/2$ \cite{ryabov/MR:2022}. In the approximate Langevin equation~\eqref{eq:Langevin-x-approx} the time-dependent stochastic force $(\muc f_y/2)\sin\left(2\phi(t)\right)\cos(\Omega t)$ competes with the deterministic force $-\dd V/\dd x$. Equation~\eqref{eq:Langevin-x-approx} corresponds to a so-called one-dimensional rocking ratchet model \cite{Reimann:2002} where the potential $V(x)$ is tilted in a stochastic way. The effective height of the barriers in $V(x)$ changes upon tilting of the potential which may produce a net particle current.

The stochastic force in the $\hat{x}$-direction depends upon the transverse component of the external force $f_y\cos(\Omega t)$, cf.\ Eq.~\eqref{eq:F}, and emerges as a result of the product $\nvec \cdot {\bm F}$ in the expression for the active speed \eqref{eq:ua}. This highlights the importance of the local detailed balance condition \eqref{eq:detailed_balance} leading to the mechanochemical coupling and hence to the force-dependent self-propulsion speed. Indeed, if we were to use a standard active Brownian particle model which neglects microscopic reversibility of chemical reactions powering the self-propulsion, the first term in the right-hand side of Eq.~\eqref{eq:Langevin-x-approx} would be missing. 

When the amplitude $(\muc f_y/2)$ of the stochastic rocking force becomes large enough, the nanoparticle starts surmounting potential barriers of $V(x)$. The asymmetry of $V(x)$, which has one slope steeper than the other, facilitates the barrier crossing in $-\hat{x}$-direction in this case. This leads to nonzero mean velocity $\langle v_x \rangle<0$ plotted in Fig.~\ref{fig:vx_vs_fy_W0}. For very large $(\muc f_y/2)$, the ratchet potential term in Eq.~\eqref{eq:Langevin-x-approx} may be neglected, such that the nanoparticle moves under the action of two stochastic forces with zero mean velocity. 

\begin{figure}[t]
\centering
\includegraphics[width=0.5\columnwidth]{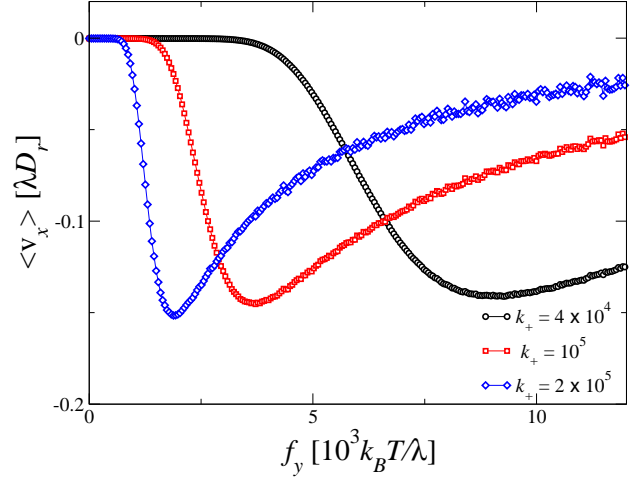}
\caption{The $x-$component of the particle average velocity, as a function of the amplitude $f_y$ of a time-dependent force $f_y\cos(\Omega t)$, which is applied perpendicular to the ratchet direction, $\Omega = \Dr$. The depth of the ratchet potential $V_0 = 10\; \kB T$. Different curves correspond to different values of the forward reaction rate constant $k_+$.}
\label{fig:vx_vs_fy_W1}
\end{figure}

If the frequency $\Omega$ of the rocking force is low compared to the inverse correlation time of the rotational diffusion, i.e., $\Omega < 4 \Dr$, the above results are influenced only quantitatively by changing $\Omega$, as is shown in Fig.~\ref{fig:vx_vs_fy_W1}. The maximal mean speed reduces roughly by half and the location of the minima slightly shifts towards larger values of $f_y$. 

\begin{figure}[]
\centering
\includegraphics[width=0.5\columnwidth]{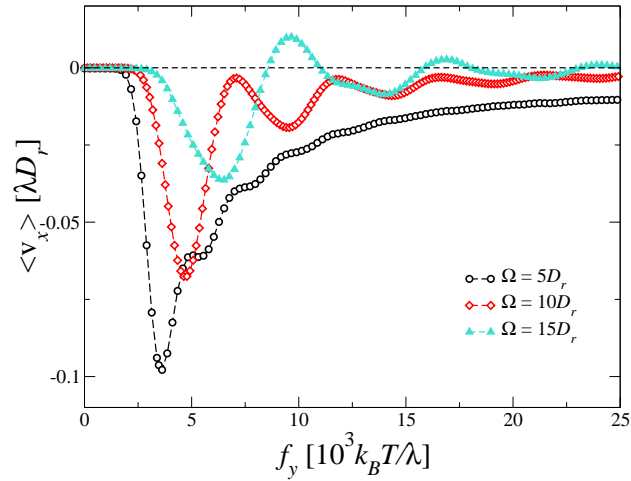}
\caption{The $x-$component of the mean velocity $\langle v_x \rangle$ as a function of the amplitude $f_y$ of a time-dependent force $f_y\cos(\Omega t)$, which is applied perpendicular to the ratchet direction. The depth of the ratchet potential $V_0 = 10\; \kB T$ and the forward reaction rate constant $k_+=10^5\;\si{s^{-1}}$. Different curves correspond to different values of the force frequency parameter $\Omega$.}
\label{fig:V10_K105_vx_diffW}
\end{figure}

\begin{figure}[t]
\centering 
\includegraphics[width=0.5\columnwidth]{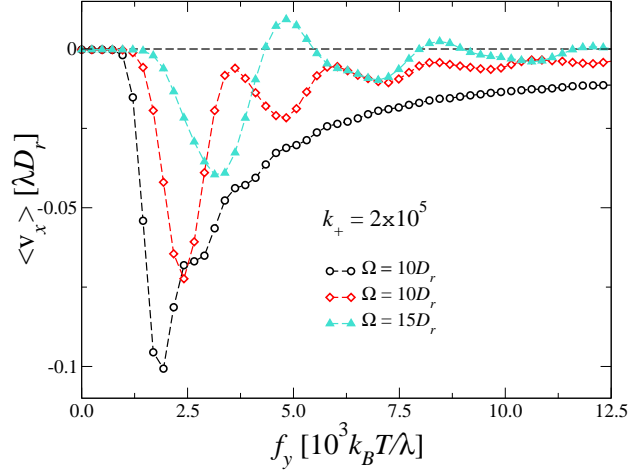}
\caption{The $x-$component of the mean velocity as a function of the frequency $\Omega$ of the transverse driving force $f_y\cos(\Omega t)$, see Eq.~\eqref{eq:F}. The forward reaction rate constant $k_+=2\times10^5\;\si{s^{-1}}$, and the depth of the ratchet potential $V_0 = 10\; \kB T$. Different curves correspond to different values of the force amplitude $f_y$. Upon rescaling the horizontal axis by a factor of 2, the data shown here collapses approximately on the ones shown in Fig.~\ref{fig:V10_K105_vx_diffW}.}
\label{fig:V10_K_2x105_vx_diffW}
\end{figure}

\begin{figure}[]
\centering
\includegraphics[width=0.5\columnwidth]{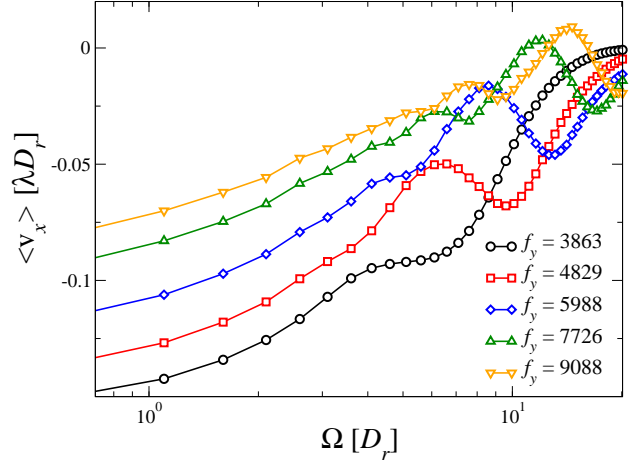}
\caption{The $x-$component of the mean velocity as a function of the frequency $\Omega$ of the transverse driving force $f_y\cos(\Omega t)$, see Eq.~\eqref{eq:F}. The forward reaction rate constant $k_+=10^5\;\si{s^{-1}}$, and the depth of the ratchet potential $V_0 = 10\; \kB T$. Different curves correspond to different values of the force amplitude $f_y$.}
\label{fig:V10_K105_vx_W_v3}
\end{figure}

Qualitatively different behaviour is observed for $\Omega \gtrsim 1/\tau_{\rm r}$, where $\tau_{\rm r}=1/4\Dr$ is the rotational diffusion correlation time. Figure~\ref{fig:V10_K105_vx_diffW} in this regime demonstrates oscillatory $\langle v_x \rangle$ as a function of $f_y$ for intermediate driving amplitudes. Surprisingly, for $\Omega$ above a certain threshold $\approx 13\Dr$ the mean velocity reverses its sign, and the nanoparticle moves along the hard ratchet direction corresponding to the $\hat{x}$-direction (magenta open triangles in Fig.~\ref{fig:V10_K105_vx_diffW}). Two representative nanoparticle trajectories $x(t)$ for the motion in the $-\hat{x}$- and $\hat{x}$-direction are presented in Fig.~\ref{fig:model}. In Fig.~\ref{fig:V10_K105_vx_W_v3} the mean velocity is plotted as a function of $\Omega$ for several values of the driving force amplitude $f_y$. Again, we observe that $\langle v_x \rangle$ oscillates with $\Omega$ provided $\Omega\gtrsim 1/\tau_{\rm r}$. The nanoswimmer speed decreases with increasing $f_y$ and the velocity reverses its sign multiple times as $f_y$ grows. The amplitude of the transverse force enters \eqref{eq:Langevin-x-approx} as $\muc f_y$, therefore we expect an approximate scaling relation to hold between $\langle v_x \rangle (f_y)$ curve obtained at different values of $\muc$, recall that $\muc\propto k_+$. We plot $\langle v_x \rangle$ as functions of $f_y$ at $k_+ = 2\times 10^5\;\si{s^{-1}}$ in Fig.~\ref{fig:V10_K_2x105_vx_diffW}, which approximately collapse on top of the curves shown in Fig.~\ref{fig:V10_K105_vx_diffW} upon rescaling the horizontal axis of Fig.~\ref{fig:V10_K_2x105_vx_diffW} by 2. The scaling is not exact, because $\muc$ also controls the effective mobility $\mueff$ and $\Deff$ in \eqref{eq:Langevin-x-approx}.

Qualitatively similar behaviour has been reported for a passive Brownian particle in a periodically rocking (in the $\hat{x}$-direction) ratchet potential \cite{Bartussek_1994}. We emphasize the following important difference between our model and that of \cite{Bartussek_1994}, while in our case the oscillating force is applied perpendicular to the ratchet  direction, i.e. in the  $\hat{y}-$direction, in \cite{Bartussek_1994} the oscillating force acts along the $\hat{x}-$direction. A passive Brownian particle does not show current rectification in the setup considered here. 

\begin{figure}[]
\centering
\includegraphics[width=0.5\columnwidth]{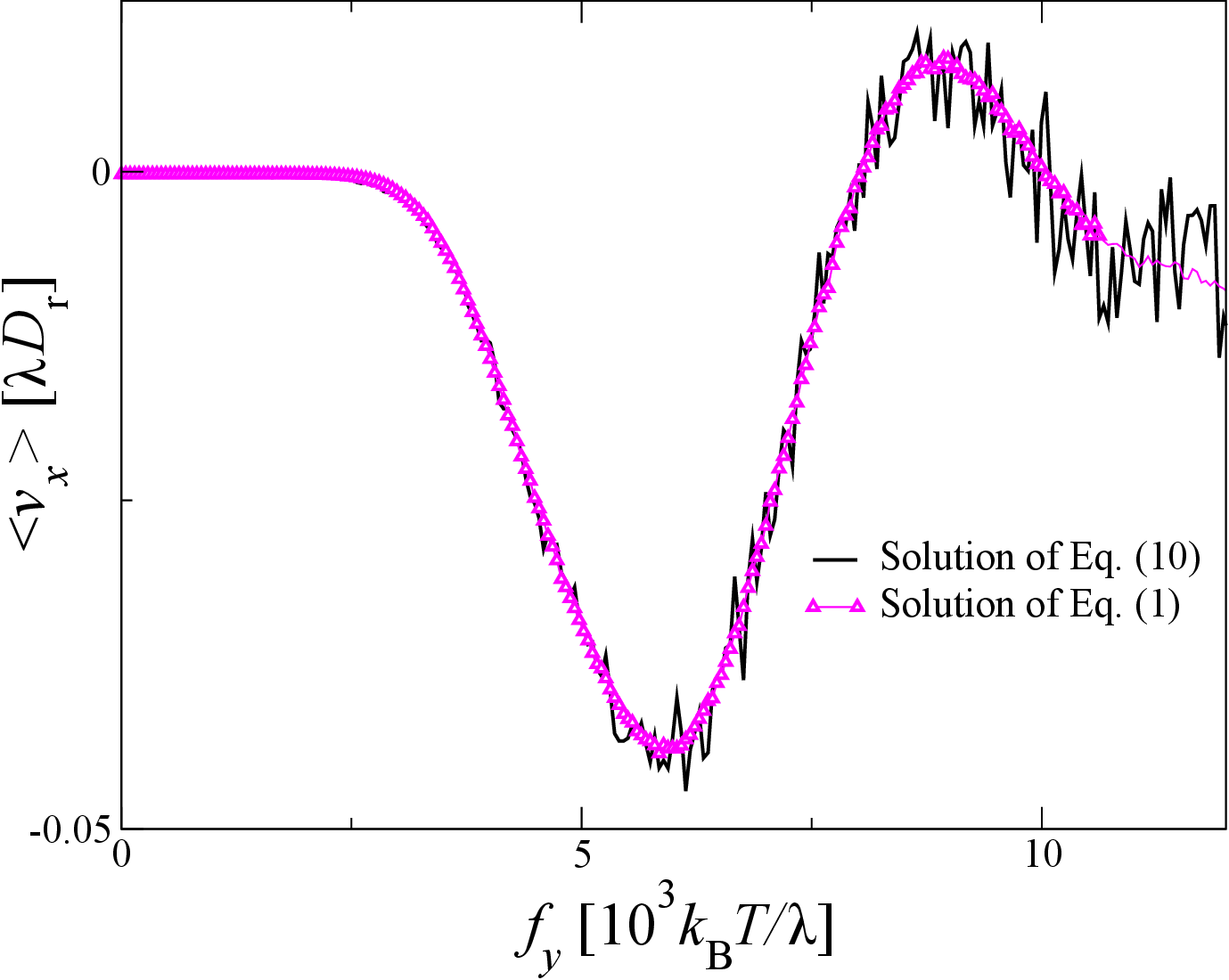}
\caption{Comparison of the predictions of the approximate model in Eq.~\eqref{eq:Langevin-x-approx}, black line, with the results of the original Langevin equation \eqref{eq:Langevin-general}, symbols. The $x-$component of the mean velocity $\langle v_x \rangle$ as a function of the amplitude $f_y$ of a time-dependent force $f_y\cos(\Omega t)$, which is applied perpendicular to the ratchet direction. The depth of the ratchet potential $V_0 = 10\; \kB T$ and the forward reaction rate constant $k_+=10^5\;\si{s^{-1}}$, $\Omega = 14 \Dr$.}
\label{fig:approx}
\end{figure}
 
This behaviour is reproduced semi-quantitatively by the approximate Langevin equation~\eqref{eq:Langevin-x-approx} where $f_y$ is replaced with $f_y \cos(\Omega t)$. We have integrated \eqref{eq:Langevin-x-approx} numerically for several representative systems and found very good agreement with the results of the full model \eqref{eq:Langevin-general}, as shown in Fig.~\ref{fig:approx}. Therefore, our system can be interpreted in terms of an effective symmetrically (with time) rocking ratchet even with time dependent transverse driving. The stochastic force $\propto \muc f_y \sin\left(2\phi(t)\right)$ acting in the $\hat{x}$-direction appears in \eqref{eq:Langevin-x-approx} as the result of the interplay of the mechanochemical coupling and the rotational diffusion, and it explains qualitatively the initial current rectification when the nanoswimmer starts drifting in the $-\hat{x}$-direction. When the amplitude $(\muc f_y/2)$ of the stochastic rocking exceeds a threshold the nanoswimmer hops predominantly over gentler sides of $V(x)$, towards $-\hat{x}$-direction. However, it fails to explain the multiple current reversals as $f_y$ grows, and the physical reason of this behaviour remains elusive.

Finally, let us note that the ratchet effect will also be present when both $ \Delta G_{\rm r} = 0 $ and $u = 0$, emphasizing again the fundamental role of the mechanochemical coupling, the second term on the right-hand side of Eq.~\eqref{eq:ua}.

\section{Summary and perspectives}

We have presented a robust mechanism for controlling the motion of chemically powered self-propelled nanoparticles moving in a fluid. The active dynamics is considered in the framework of a Markov jump process where the rates of the forward and backward reactions obey the local detailed balance condition. A coarse-grained description (diffusion approximation) of the process results in the stochastic expression \eqref{eq:ua} for the self-propulsion speed, which bears an explicit dependence upon external forces as a consequence of the mechanochemical coupling. 

Within such a thermodynamically consistent description, the mechanochemical coupling leads to the rectification effect for the active dynamics across a ratchet potential $V(x)$. While the time-independent potential $V(x)$ is asymmetric and periodic along the $x$-axis, the ratchet effect is triggered by a time-periodic and space-independent transverse force acting in the $\hat{y}$-direction. The transverse force induces a directional particle motion along the $x$-axis. 

For the values of model parameters corresponding to active nanoparticles, the dynamics along the $x$-axis can be approximately described by an equation~\eqref{eq:Langevin-x-approx}, which takes the form of a Langevin equation for a one-dimensional rocking ratchet. This approximate description contains a time-dependent rocking force in the $\hat{x}$-direction even though no actual time-dependent forces are acting in the $\hat{x}$-direction in the original model. The effective force is mediated by the mechanochemical coupling and has a stochastic amplitude modulated due to the rotational diffusion of the nanoswimmer. When the frequency of the transverse force equals to or is larger than the inverse of the rotational diffusion correlation time, the oscillatory behaviour of the mean velocity emerges. In particular, we observe multiple velocity reversals as a function of the amplitude and the frequency of the transverse force.

Relatively large absolute values of amplitude of the transverse driving required for the rectification of the nanoswimmer current reported here can be rationalised as follows. In the original microscopic jump-diffusion model with the exponential dependence of the reaction rates \eqref{eq:detailed_balance} on the external force, the threshold amplitude of the force needed to rectify the nanoswimmer velocity, would be orders of magnitude smaller. The large values of the rectifying force reported here result from passing to a continuum limit \eqref{eq:ua} when the exponential dependence of the rates on the force is replaced by a linear one. 

The presented results highlight the importance of the fundamental principle of microscopic reversibility for chemically-powered dynamics at the nanoscale, contrary to the active motion of micron-sized colloids, where MR of their self-propulsion mechanism can be neglected, due to the relatively large dimensions of the particles. The ratchet effect described here is absent for passive Brownian particles or in the active Brownian particle models without the mechanochemical coupling.

The reported phenomena can have two general implications. For nanoswimmers, processes responsible for the self-propulsion can be explored indirectly only. Due to the small dimensions, there are no experimental methods capable of directly observing the reactions and ambient fluid flows that cause the particle movement. In this regard, our general aim was to propose an observable effect, which would solely rely on the minimal assumption (MR) regarding the self-propulsion mechanism. (i) Such an effect can be used to test physical principles inducing self-propulsion at the nanoscale. For example, it might be useful in resolving a longstanding debate regarding the self-propulsion of active enzymes. (ii) It can inspire the development of new methods of control of nanoswimmers at the single-molecule level. This can be crucial for targeted cargo delivery at the nanoscale, which has not been achieved experimentally yet.

\section*{Acknowledgements}
We acknowledge financial support from the Portuguese Foundation for Science and Technology (FCT) under Contracts nos.\ PTDC/FIS-MAC/5689/2020, UIDB/00618/2020, and UIDP/00618/2020 and from the Department of Physics and Mathematics at Nottingham Trent University (grant no.\ 01/PHY/-/X1175). AR gratefully acknowledges financial support from the Czech Science Foundation (project no.\ 20-24748J). Computational resources were provided by the e-INFRA CZ project (ID:90140), supported by the Ministry of Education, Youth and Sports of the Czech Republic. 

\section*{References}
%
\end{document}